# Interoperability between heterogeneous federation architectures: illustration with SAML and WS-Federation

Mikaël Ates
*mikael.ates@univ-st-etienne.fr*

Christophe Gravier
*christophe.gravier@univ-st-etienne.fr*

Jeremy Lardon
*jeremy.lardon@univ-st-etienne.fr*

Jacques Fayolle
*jacques.fayolle@univ-st-etienne.fr*

Bruno Sauviac
*bruno.sauviac@univ-st-etienne.fr*

*SATIn Team – DIOM Laboratory
Institut Supérieur des Techniques Avancées de Saint-Etienne
23 rue Paul Michelon 42023 Saint-Etienne France*

## Abstract

*Digital identity management intra and inter information systems, and, service oriented architectures security, are the roots of identity federation. This kind of security architectures aims at enabling information systems interoperability.*

*Existing architectures, however, do not consider interoperability of heterogeneous federation architectures.*

*In this paper, we try to initiate an in-depth reflection on this issue, through the comparison of two main federation architecture specifications: SAML and WS-Federation.*

*Firstly, we propose an overall outline of identity federation. We furthermore address the issue of different federation architectures interoperability. Afterwards, we compare SAML2 and WS-Federation1.1B. Eventually, we give a way to make them interoperate.*

## 1. Introduction

Information systems interconnection is a favored way to share digital resources between institutional or business partners. One of the main point of this interconnection is how to control resource access thereby implying the identity management systems interoperability. To a large extend, identity federation offers a solution, opening the way to connect information systems security domains.

Two architectures emerged from web technologies through successive evolutions, and became major identity federation architectures: SAML and WS-Federation.

In this paper, we address the case of the interoperability between these two architectures. Firstly, we propose an overall outline of identity federation. Then, we discuss about the interconnection of identity federation architectures coming from heterogeneous specifications. Afterwards, we compare SAML2 and WS-Federation1.1B. Eventually, we define the ways of convergence, and therefore, of interoperability.

## 2. Trust outline

Identity federation consists in defining an identity information transport architecture to retrieve the clearance of access from the security domain the requesting digital identity belongs to. This is based on trust architectures. Their origins come from three issues:

- The multitude of administrative processes within identity management architectures (Authentication, Authorization, Accounting and Auditing AAAA).
- Identity management between information systems.
- Web service oriented architectures security.

## 2.1. Administrative task delegation

The situation, where each application is in charge of processing every administrative task of identity management, creates many factors that potentially compromise the information system security. One of them is the multitude of authentication processes requiring each time credential reemissions over networks. Single sign-on architectures based on authentication delegation to trusted third parties overcome this issue [1][2]. The authentication delegation principle can be first attributed to the Kerberos architecture[3]. This type of architectures relies on trust links establishment between service providers and administrative authorities. Trust linked entities form a circle of trust at the heart of which service providers accept claims made by administrative authorities. In practice, it implies secrets shared among entities aiming at signing their assertions, which often relies on public key infrastructure.

## 2.2. The opening of information systems

The opening of information systems makes internal resources available to external parties, human beings or standalone applications. Administrative load of multiple identity management systems, and related financial costs, prevent information systems to directly include identity management systems of external parties. In other words, the main point is to use existing identities, so far as obtaining the most relevant information, rather than creating new identities for each application. As a matter of fact, this opening must go hand in hand with the opening of identity management systems and their interconnection. Trust links establishment between information systems also overcomes this issue.

Generally, each information system belongs to a distinct organizational entity (firm, institution, etc...). Thus, the concept of federation makes sense for information systems interconnection. In a first time, various entities formalize their partnership to make up a federation. They therefore define their common identity management policies which are afterwards carried out into the federation architecture.

## 2.3. Service oriented architectures security

Software interconnection inside or between information systems, also as the conception of software spread over multiple information systems, cause complex trouble for identity management. It means to ensure secured exchanges and to manage identities of standalone applications at the same time. Identity federation architectures are service oriented architectures which is an identity management security layer bound to third service oriented architectures. Hence, applicative messages exchanges depend on decisions taken at this security layer and are secured through security informations inserted in their headers at this security layer too.

# 3. Federation architecture

This section deals with the overall aspects of federation architectures.

## 3.1. Security information

The three points defined in section 2 rely on the circulation of security informations about identities inside or across security domains. These informations can be:
- descriptives (credentials and attributes)
- resulting of identity administrative processes.

## 3.2. Trust architectures and information security lifecycle

Trust architectures rely on trust links establishment based on shared secrets bound to sign security informations. The trust architectures are in charge of defining the lifecycle of the security informations. The trust architectures allow the conception of various trust topologies based on direct and indirect trust links and ensure messages security.

## 3.3. Identity federation: entity roles and profiles

Federation architectures rely on trust architectures in order to transport and manage security informations. They therefore extend those architectures for defining the federation functions, hence administrative tasks delegation and how the security informations must be employed. The federation architectures define the roles of the entities taking part into federation. Finally, a federation architecture must provide a way to be deployed using existing applicative transport protocols.

## 3.4. Clients

The federation architectures have to take into account the different interactions of a user or a standalone application with the identity management layer. It means two types of clients:
- web browsers, passive relays of applicative messages;
- enhanced clients, user interfaces or standalone applications, being able to directly interact with the entities members of the federation architecture, that is to say, being able to consume identity web services.

## 4. Interoperability of heterogeneous federation architectures

The federation architectures enables services for identities of a third security domain. The service providers consume informations provided by authorities with which they share direct or indirect trust links. As a consequence, the main point treated in this paper is how to make a service provider consume security informations provided by an authority issued from a different specification. In this section, we care about the entity in charge of the interoperability processes.

If service providers and authorities of different specifications are directly connected through trust links, they will be responsible for the interoperability processes. Two possibilities:
- If the federation architecture topology is authority-centered, the authority is in charge of ensuring interoperability (cf. fig1).

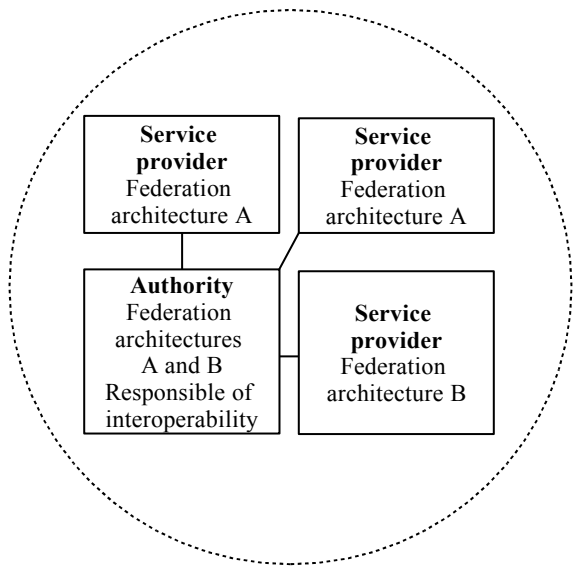

— trust link
····· circle of trust

*fig 1. Responsibility of interoperability in a federation architecture topology authority-centered with direct trust links.*

- If the federation architecture topology is service provider-centered, the service provider is in charge of ensuring interoperability (*cf. fig2*).

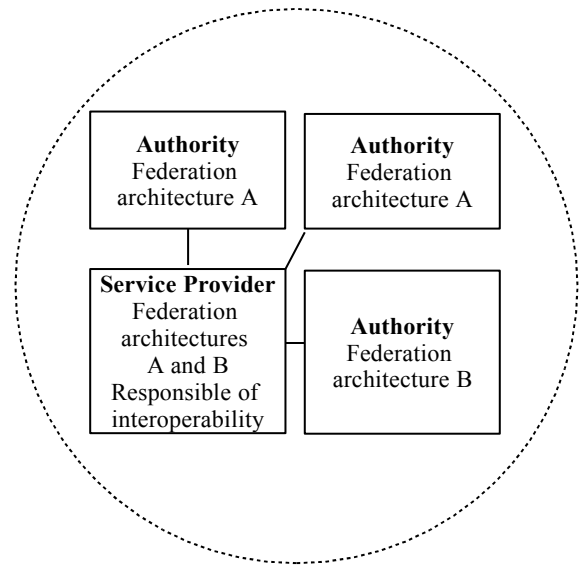

— trust link
····· circle of trust

*fig 2. Responsibility of interoperability in a federation architecture topology service provider-centered with direct trust links.*

Nevertheless, in more complex topologies, where circles of trust include multiple authorities and service providers, it is interesting to assign the responsibility of the interoperability processes to a dedicated third party. Entities issued from different specifications can interoperate without being modified thanks to the third party. As a matter of fact, they must be indirectly trust linked by the dedicated third party (cf. fig3).

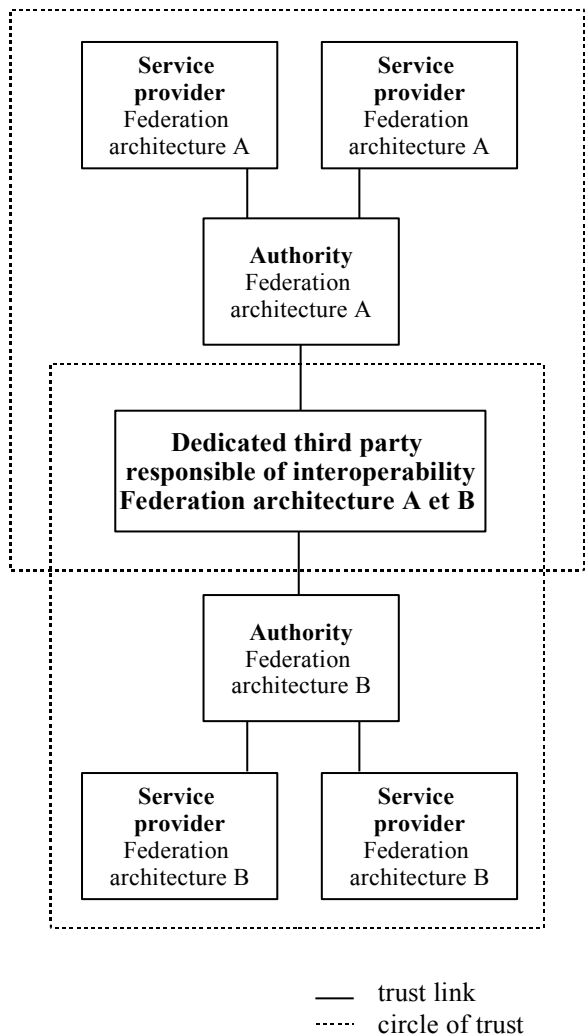

*fig 3. Responsibility of interoperability in a complex federation architecture topology allowing indirect trust links.*

## 5. Interoperability between SAML2 et WS-Federation1.1B

We illustrate the description made in the previous section with a study on the SAML2[4] and the WS-Federation1.1B[5][6] specifications. The purpose is to determine the necessary interoperability processes and the ways for implementing the dedicated third party previously described. In other words, the goal of this section is to define how a SAML2 service provider can obtain and consume a security information issued by a WS-Federation 1.1B authority. And inversely, with a WS-Federation service provider and a SAML2 authority. This section is organized as follow. Firstly, we present the specifications. Next, we compare SAML2 and WS-Federation1.1B. Finally, we propose our solution to make them interoperate.

### 5.1. Specifications outline

The OASIS[1] consortium, with the SAML norm, and the Liberty Alliance consortium, with the ID-FF[2] spécifications based on SAML 1.1, worked on closed federation architecture which resulted in the proposal of SAML2.

The WS-I[3] consortium developed a federation architecture named WS-Federation which relies on other specifications, mainly WS-Trust[7] and WS-Security[8], both normalized by the OASIS.

### 5.2. Comparison between SAML2 et WS-Federation1.1B

#### 5.2.1. Technical comparison

This comparison is provided in Tab1 for readability purpose.

#### 5.2.2. Discussion

In the rest of this paper, we will refer to the authority producer of security informations as Identity Provider (henceforth IP) and to the security informations consumer as Service Provider (henceforth SP).

SAML2 was initially conceived to answer to the single sign-on issue within a federation architecture. Implied applications were mainly web ones. That is the reason why the architecture was designed for a passive client (a web browser). The protocol messages exchanges are always triggered by the SP when it needs to establish a security context. SAML2 provides a baseline set of profiles for the use of assertions and protocols to accomplish specific uses cases. The protocols used imply which statements will be contained into the assertions. The web browser as a user interface cause the transport to be based on HTTP 1.1, and lead the client to be a passive relay of protocol messages. Direct protocol messages exchanges between an asserting party and a relying one are done in SOAP.

SAML provides an enhanced client profile (SAML ECP) which defines how a web active client should take part in the SAML message protocol exchanges. The SP is still the trigger of the protocol messages exchanges. Hence, when the client requests a service access and the SP requires a security context, it returns a SOAP request in the HTTP response (« PAOS binding/protocol »). The client is then in charge of treating the SOAP request containing an SAML authentication request signed by the SP. The SP gives in the SOAP header his list of trusted IP. The client selects one and submits it the authentication request with a new SOAP request. If the selected IP has not already established a security context, or the SP forces the reauthentication, the client authenticates against the IP. The IP delivers the assertion containing the authentication statement. The client returns the assertion to the SP in a SOAP response in a HTTP request (« PAOS binding/protocol »). The SP establishes the security context for the user and delivers the resource initially requested.

---
[1] Advancing Open Standards for the Information Society
[2] IDentity-Federation Framework
[3] Web Services Interoperability

The WS-Federation trust architecture is based on WS-Trust specifications. WS-Trust was initially conceived for the security of web services oriented architectures. WS-Trust provides an architecture for an active client, the requestor, actor of the protocol messages exchanges. It is in charge of triggering these exchanges. The requestor can be an enhanced web browser, as a user interface, or a standalone web service client. WS-Trust defines a protocol of exchanges between requestors and security token servers allowing to manage the token lifecycle. WS-Trust relies on WS-Security specifications to ensure protocol message security and to provide a way to describe security informations. WS-Federation defines various profiles of STS according to the type of tokens they are able to deliver: Authentication, Autorization, Attributes and Pseudonyms. It is important to notice that the WS-Trust protocol, unlike SAML, is generalist and does not provide specialized requests according to the types of tokens required. WS-Federation also provides a model where each web service uses its own WSDL file. It defines the service federation capabilities and the security requirements through the integration of a document named policy respecting the WS-SecurityPolicy format. The federation capabilities are named federation metadatas and describe the SOAP endpoint on the web service. The requirements defined by the policy indicate the claims the requestor will have to present in his requests to obtain the service access. Reciprocally, the requestor deduces from the policy the claims it has to obtain from the IP/STS.

WS-Federation provides a passive client profile. Initially, the requestor was in charge of triggering the WS-Trust protocol messages exchanges. The passive client profile lets the SP trigger the exchanges, like in SAML. Not only the SP must be aware of the claims it requires, but it also becomes the requestor. Which consequently leads to the definition of a protocol with specialized requests.

The SAML request relaying, named SAML proxying, and the token exchange in WS-Federation, are the basis for the deployment of trust architectures made of indirect trust links between producers of security informations and their consumers. Indirect trust links allows the IP chaining and hence, hierarchical trust architectures. The third party responsible of the interoperability processes described in this paper relies on this capability.

|  |  | SAML2 | WS-Federation1.1.B |
|---|---|---|---|
| Semantic | Authority producer of security informations (IP) | Asserting party or identity provider | Secutity token server or identity provider |
|  | Security informations consumer (SP) | Assertion consumer or relying party or service provider | Relying party or ressource provider |
|  | Subject of the security information | Principal | Principal |
|  | Passive client | Web browser | Web browser requestor |
|  | Active client | Enhanced client or enhanced proxy | Web service requestor |
| Security Information | Label | Assertion | Security token |
|  | Specifications | Assertions and protocols for the OASIS SAML v2. | Web Services Security: SOAP Message Security1.1 (WS-Security 2004) |
|  | XML schema namespace | urn:oasis:names:tc:SAML:2.0:assertion | http://docs.oasis-open.org/wss/2004/01/oasis-200401-wss-wssecurity-secext-1.0.xsd |
|  | Description | Contains statements made by an Issuer about a Subject. Three kinds of assertion statements:<br>. authentication,<br>. autorization decisions,<br>. attributes. | Three kinds of security tokens:<br>. « username »,<br>. binary,<br>. XML (e.g: SAML assertion).<br>Contains a collection of « claims ». |
| Provisionning of federation metadatas (used to diffuse public keys and to describe endpoint references of  federation services) | | Dynamic | Dynamic with the WS-MetadataExchange protocol [9] |
| Pseudonimity[10][11] | | It exists two formats of identifiers ensuring privacy:<br>. The transient identifier is a temporary pair-wise pseudonym.  It usually changes at each user session and is different for each link between IP and SP.<br>. The persistent identifier is a persistent pair-wise pseudonym different for each link between IP and SP. As the transient one, it prevents the discovery of the subject's identity or activity, but it allows account linking (aka identity mapping). | WS-Federation offers two equivalent pseudonym mecanisms:<br>. The pair-wise identifier, equivalent as the persistent one of SAML.<br>. A transient identifier coupled with a pseudonym registring service allowing the account linking |
| Protocol messages | Specifications | Assertions and protocols for the OASIS SAML v2.0 | WS-Trust 1.3 |
|  | XML schema namespace | urn:oasis:names:tc:SAML:2.0:protocol | http://docs.oasis-open.org/ws-sx/ws-trust/200512 |
|  | Request type | The request type depends on the security information type expected. In other words, the request type employed depends on the profile used. Inversely, the statements contained in assertion depend on the request type employed. | WS-Trust provides a single type of request. The claims contained in security token depends on the claims required by the SP. The requirements of the SP are defined in a policy. The policy format must respect  the WS-SecurityPolicy[12] format. |
|  | Trigger of protocol messages exchanges / Passive client | SP | SP. Two possibilities to the SP for requesting the IP:<br>. With a set of URL parameters.<br>. With a WS-Trust request. |
|  | Trigger of protocol messages exchanges / Active client | SP | Active client |
| Authorization | | Authorizations management is not included in SAML anymore. XACML issued by the OASIS provides an authorization architecture. The XACML authorization statements can be conveyed in attributes statements of SAML assertions as it is defines in the XACML attribute profile. | WS-SecurityPolicy offers a way to define policy for each entity of a WS-Federation architecture. It constitutes the authorization architecture of WS-Federation. |
| Trust architecture | | An IP can act as an authentication request relay. In other words, if it can not satisfy a request, it can request by itself another IP to satisfy the initial request and acts as an assertion consumer, as an SP usually acts. In fact, it keeps the assertion as is and resigns it. This mechanism has been conceived for a passive client but works as well for an active client. The IP returns a SOAP request instead of a SOAP response, and the client submit the SOAP request to another IP. | Once the client has determined the claims required by the SP and if  the IP implied is different from the IP affiliated to the SP, it obtains the token from the IP and exchanges it to the IP affiliated to the ressource. With the passive client profile,like with SAML, the SP redirects the web browser to its IP which redirect to the IP deduced from the principal. |

## 5.3. Interoperability and passive client

The interoperability processes are fundamentally the same for a passive or an active client. Therefore, we will illustrate our proposition with an architecture based on a passive client.

### 5.3.1 Existing works

At the present day, we did not found any full implementations of the passive or active profiles of WS-Federation1.1B. The ADFSv1 on Windows Server 2003 R2 is an implementation of the interoperability profile which is a subset of the passive profile of WS-Federation1.0. Works were conducted to make it interoperate with Shibboleth v1.3f. based on SAML1.1[13]. These works lead to the elaboration of a Shibboleth extension module. The extension module is installed directly on the SP or IP which needs interoperability. The IP proxying is not supported by Shibboleth v1.3f which inhibits the possibility of implementating a third party dedicated to the interoperability.

This interoperability architecture is based on the conversion of the SAML requests in the WS-Federation passive requestor interoperability profile equivalent requests. More precisely, the XML documents corresponding to the SAML requests and responses are respectively converted in URL parameters and in a *<wst:RequestSecurityTokenResponse>* WS-Trust document.

### 5.3.2 Principle

The WS-Federation passive requestor profile specifications provide two ways for the SP to request the IP:
- Through a set of URL parameters. The way exploited in the interoperability ADFS/Shibboleth previously described.
- Through a WS-Trust request document given by the SP to the IP into a URL parameter.

The second way seems to be the most interesting for us in the perspective of implementing the interoperability of the WS-Federation passive profile with the SAML profiles in a dedicated third party.

Hence, the main task will result in translating:
- A WS-Trust *<wst:RequestSecurityToken>* document in SAML request documents, for example *<samlp:AuthnRequest>*, and inversely.
- A WS-Trust *<wst:RequestSecurityTokenResponse>* document in a SAML *<samlp:Response>*, and inversely.

The third party dedicated to the interoperability will be in charge of performing these transformations and to resign them to establish the indirect trust link. We illustrate this principle for the authentication delegation profiles.

### 5.3.3. Transport

In both WS-Federation and SAML, the client is redirected by the SP to an IP with a HTTP 302 error. The authentication request is transmitted by the following URL parameters:
- For SAML: parameter *SAMLRequest* containing an XML document of *<samlp:AuthnRequest>* type.
- For WS-Federation: parameter *wa* containing the *wsignin1.0* value and parameter *wreq* containing an XML document of *<wst:SecurityTokenRequest>* WS-Trust type.

A valid authentication response is conveyed by a HTTP POST with the following parameters:
- For SAML: parameter *SAMLResponse* containing an SAML assertion.
- For WS-Federation: parameter *wresult* contaning an XML document of *<wst:SecurityTokenRequestResponse>* WS-Trust type which itself contains an SAML assertion.

### 5.3.4. The request

The objective here is to convert a SAML request in a WS-Trust request.

#### 5.3.4.1. SAML « AuthnRequest »

As we mentioned in 5.2., WS-Trust requests are generalist. Therefore, it is necessary to translate in WS-Trust the fact that the information expected as a response for a *<samlp:AuthnRequest>* SAML request must be the result of an authentication process.

Hence, the *<samlp:AuthnRequest>* SAML request is converted in the *<wst:RequestSecurityToken>* WS-Trust request containing the following element:

*<wst:RequestType>http://docs.oasis-open.org/ws-sx/ws-trust/200512/Issue</wst:RequestType>*

Furthermore, the token type expected for the response is an assertion. The request must also contain the following element:

*<wst:TokenType>* urn:oasis:names:tc:SAML: 2.0:assertion *</wst:TokenType>*.

### 5.3.4.2. The subject name

A *<samlp:AuthnRequest>* SAML request contains the element *<samlp:NameIDPolicy>* indicating the type of subject name expected in the response. It can be translated by a request of an authorization claim of the same type. For example with an email address, we obtain these two requests:

- *<samlp:authnRequest>*:

*<samlp:NameIDPolicy Format="urn:oasis:names:tc:SAML:1.1:nameid-format:emailAddress" </samlp:NameIDPolicy>*

- *<wst:SecurityTokenRequest>*:

*<wst:Claims Dialect= http://schemas.xmlsoap.org/ws/2006/12/authorization/authclaims ><auth:ClaimType Uri="urn:oasis:names:tc:SAML:1.1:nameid-format:emailAddress" /> </wst:Claims>*

### 5.3.4.3. Authentication level

An SAML SP indicates within an authentication request the authentication context expected using the *<samlp:RequestedAuthnContext>* element. SAML provides 25 authentication context XML schemas.

In a WS-Trust request, it is done with a *<wst:authenticationType>* element for which a set of values are defined by WS-Federation.

The WS-Federation defined values are less numerous than the SAML authentication context schemas. However, a WS-Federation implementation can take in charge some of the SAML authentication context schema to ensure the interoperability for the missing WS-Federation authentication context.

### 5.3.5. The response

The objective is to translate a SAML response in a WS-Trust response.

The token type expected in the *<wst:SecurityTokenRequestResponse>* WS-Trust response is a SAML assertion. The third party will have to extract, to resign and to reformat it in a *<samlp:Response>* SAML response.

### 5.3.6. Generalization

We have settled for explaining the principle on which we based the interoperability between SAML2 and WS-Federation1.1B. To specify precisely this interoperability, it is necessary to apply this principle to convert each SAML requests and responses in WS-Trust requests and responses and inversely. Many more parameters than those previously cited have to be taken in account, eg, parameters about the information security lifetime.

To provide attributes in cross-federation, it is also necessary to define a common namespace. WS-Federation offers a very limited claim namespace based on the non-normative document « Passive Requestor Interoperability Profile » [14]. SAML provides five attributes profiles: basic, X500/LDAP, UUID, DCE PAC et XACML.

For ensuring privacy, it needs to make interoperate the pseudonyms systems, ie, the dedicated STS for WS-Federation, and the use of transcient and persistent pseudonyms for SAML.

Finally, it does not exist a dynamic metadatas discovering service in SAML. It can be considered as the least common denominator and thus, the trust links will have to be pre-established with the third party.

The interoperability with active clients profiles is based on the same principle except that the client is also active in the interoperability processes. Not because it plays a role in XML document and protocol messages conversions, but it has indeed to support the transport protocols of both WS-Federation and SAML ECP profile specifications.

## 6. Conclusion

The federation architectures are still evolving. Nevertheless, overall trends emerge and converge to the interoperability of heterogeneous architectures.

The « Identity 2.0 » philosophy places the user/consumer at the heart of the exchanges concerning his identity. This perfectly corresponds to active clients federation architectures: the requestor of WS-Trust, the SAML ECP profile, and the Active Client in the Liberty Alliance ID-WSF2.0 specifications. A main feature of the active client is to provide attributes by his own. It is also a better experience for a user in the management of his identities. The result of the exposed work leads to think that active clients are the main convergence point of the federation architectures.

From the basis settled in this paper, our future works could be to propose a model of interoperability between SAML and WS-Federation and an associated implementation.